\documentclass{article}
\usepackage{spconf,amsmath,graphicx,url,subfig, graphicx, cite,booktabs}
\usepackage{amsfonts,amssymb}
\usepackage{xcolor}

\title{DualVC 2: Dynamic Masked Convolution for Unified Streaming and Non-Streaming Voice Conversion}

\name{
\begin{tabular}{c}
Ziqian Ning$^{1,2}$, Yuepeng Jiang$^{1}$, Pengcheng Zhu$^{2}$, Shuai Wang$^{3}$, Jixun Yao$^{1}$, Lei Xie$^{1*}$, Mengxiao Bi$^{2}$\thanks{* Corresponding author.}
\end{tabular}
}
\address{
   $^1$Audio, Speech and Language Processing Group (ASLP@NPU), School of Computer Science, \\ Northwestern Polytechnical University, Xi'an, China\\
  $^2$Fuxi AI Lab, NetEase Inc., Hangzhou, China\\
  $^3$Shenzhen Research Institute of Big Data, \\The Chinese University of Hong Kong, Shenzhen (CUHK-Shenzhen), China
  }

\begin{document}
\ninept
\maketitle
\begin{abstract}
Voice conversion is becoming increasingly popular, and a growing number of application scenarios require models with streaming inference capabilities. The recently proposed DualVC attempts to achieve this objective through streaming model architecture design and intra-model knowledge distillation along with hybrid predictive coding to compensate for the lack of future information. 
However, DualVC encounters several problems that limit its performance. First, the autoregressive decoder has error accumulation in its nature and limits the inference speed as well. Second, the causal convolution enables streaming capability but cannot sufficiently use future information within chunks. Third, the model is unable to effectively address the noise in the unvoiced segments, lowering the sound quality.
In this paper, we propose DualVC 2 to address these issues. Specifically, the model backbone is migrated to a Conformer-based architecture, empowering parallel inference. Causal convolution is replaced by non-causal convolution with a dynamic chunk mask to make better use of within-chunk future information. Also, quiet attention is introduced to enhance the model's noise robustness.
Experiments show that DualVC 2 outperforms DualVC and other baseline systems in both subjective and objective metrics, with only 186.4 ms latency. Our audio samples are made publicly available\footnote{Demo: https://dualvc.github.io/dualvc2/}.

\end{abstract}

%
\section{Introduction}
\label{sec:intro}

Voice conversion (VC) aims to convert the voice from one speaker to another while maintaining the same linguistic content~\cite{vcoverview}. VC has a wide range of application scenarios, such as movie dubbing~\cite{bgmvc,expressive-vc}, privacy protection~\cite{voiceprivacy2022}, and communication aids for speech-impaired people~\cite{ELECTROLARYNGEAL}. Recently, VC has also become more popular in the field of real-time communication (RTC), such as live streaming, online meetings, and voice chat in online gaming. These applications require VC models to have streaming capabilities.

Classical VC models~\cite{acevc,vqmivc,stylevc} operate at the utterance level, converting entire utterances into the desired target speaker timbre. While showing remarkable naturalness and high expressiveness, these non-streaming models cannot be applied to real-time applications. On the contrary, streaming VC models~\cite{dualvc,ibfvc,DBLP:conf/interspeech/YangDYZX22,fasts2svc, parrotron} have the ability to process input in real-time, either frame-by-frame or in chunks with multiple frames. However, due to the absence of future information during streaming inference, they still lag behind their non-streaming counterparts, exhibiting relatively lower intelligibility, poorer sound quality, and inferior speaker similarity.


In our previous work~\cite{dualvc}, DualVC has attempted to address these problems by employing a combination of intra-model distillation and hybrid predictive coding (HPC). All convolutional layers in the base model are replaced by dual-mode convolution blocks, each comprising two parallel basic convolutional layers, one causal for streaming mode and the other non-causal for non-streaming mode. 
A knowledge distillation loss is calculated between the encoder output of the two modes to pull the hidden representation of the streaming mode close to the non-streaming mode, enhancing the performance of the streaming mode. On the other hand, HPC combines the advantages of contrastive predictive coding~\cite{cpc} and autoregressive predictive coding~\cite{apc}. The common feature structure captured by using the HPC module allows the model to infer future information to some extent. 
Despite its effectiveness, DualVC faces several problems that limit its performance. First, the autoregressive decoder is limited to frame-by-frame decoding and cannot be parallelized, resulting in increased latency. Also, autoregressive generation of spectrogram leads to error accumulation, causing a gradual decline in conversion quality. Second, in the chunk-based streaming inference, pure causal convolution fails to fully exploit future information within the current chunk. Third, background noise in unvoiced frames cannot be properly removed and can leak into the output.
\begin{figure*}[!htbp]
\centering
\includegraphics[scale=0.45]{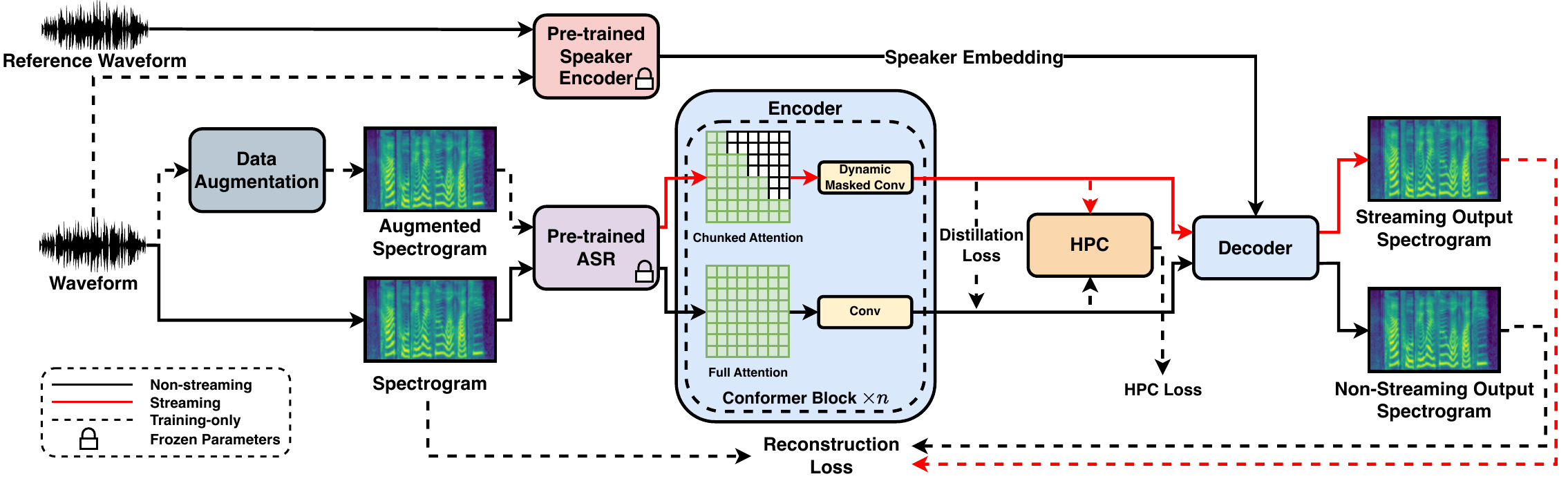}
\vspace{-6pt}
\caption{The architecture of DualVC 2.}
\label{fig:model}
\vspace{-12pt}
\end{figure*}

In this paper, we present DualVC 2, an efficient streaming voice conversion model designed to deliver faster speed and better stability, which can be applied in both streaming and non-streaming scenarios. Based on the popular recognition-synthesis framework~\cite{recognition-synthesis}, DualVC 2's backbone is built on Conformer~\cite{Conformer} blocks, leveraging its remarkable ability to capture contextual information and facilitate parallel inference. Following the concept of dynamic chunk training as proposed in WeNet~\cite{wenet}, the model can be applied to different chunk sizes to meet the needs of different latencies.
Unlike previous streaming VC models~\cite{dualvc,ibfvc,DBLP:conf/interspeech/YangDYZX22,fasts2svc, parrotron} that relied on causal convolutions for continuous inference without access to the future information, DualVC 2 uses classical non-causal convolutions that make better use of within-chunk future context and address potential feature discontinuities that cause clicking sounds between adjacent chunks through dynamic masked convolutions. In addition, to strengthen the model's robustness to noise, we incorporate a quiet attention mechanism\footnote{https://www.evanmiller.org/attention-is-off-by-one.html\label{quiet-attention}}.
Notably, our approach also integrates the HPC module and intra-model distillation proposed in our previous work~\cite{dualvc}. Data augmentation is also adopted to further increase the noise robustness and intelligibility of the model. 
Through extensive experiments, DualVC 2 shows superior conversion quality over DualVC~\cite{dualvc} and IBF-VC~\cite{ibfvc}. Compared to DualVC, the inference speed of DualVC 2 is increased by about 70\%, with an RTF of only 0.165, while the latency of the entire pipeline is reduced from 252.8 ms to 186.4 ms on a single-core CPU. 
\label{sec:format}

\section{Proposed Approach}

As illustrated in Fig.\ref{fig:model}, DualVC 2 is built on the popular recognition-synthesis framework, comprising an encoder, a decoder, and an HPC module~\cite{dualvc}. Initially, the encoder of a pre-trained streaming automatic speech recognition (ASR) model extracts bottleneck features (BNFs) from the input spectrogram. These BNFs are then forwarded to the encoder to further extract contextual information. The HPC module, which is only used during the training phase, facilitates the encoder to extract more effective latent representations through unsupervised learning methods. Subsequently, the target speaker embedding extracted by a pre-trained speaker encoder model is concatenated to the latent representation and provided as input to the decoder. Finally, the decoder generates the converted spectrogram with the target speaker timbre.

\subsection{Streamable Architecture}

As a dual-mode model, DualVC 2 is able to perform conversions both with full context in non-streaming mode and with limited context in streaming mode. To accomplish this, we adopt dynamic chunk training (DCT), which is proposed in ~\cite{wenet}. The DCT idea involves varying the chunk size dynamically by applying a dynamic chunk mask to the attention score matrix for each self-attention layer. During training, there is a 50\% chance of using the full sequence and in the rest of the cases, the chunk size is randomized between 1 (= 12.5 ms) and 20 (= 250 ms). Following the setup in our previous work, dual-mode convolution is also applied in DualVC 2 which consists of two parallel basic convolution layers for streaming and non-streaming modes respectively. Different from DualVC, the causal convolution layer is replaced by non-causal convolution with dynamic masks, which will be introduced in the next section. In line with DCT, dual-mode convolution is set to non-streaming mode when using full sequence inputs and to streaming mode when using random chunk inputs.

\subsection{Dynamic Masked Convolution}
\begin{figure}[ht]
 
\centering
\includegraphics[scale=0.7]{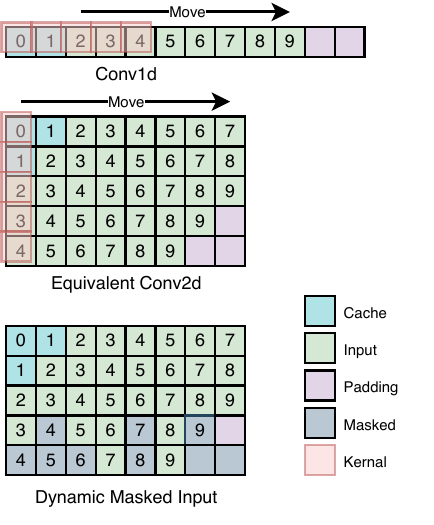}
\vspace{-8pt}
\caption{Dynamic Masked Convolution Module. Non-causal 1D convolution is replaced by equivalent 2D convolution. A dynamic mask is applied to simulate different lengths of future frames in the convolutional kernel receptive field.}
\label{fig:conv}
\vspace{-6pt}
\end{figure}

Streaming models frequently adopt causal convolutions~\cite{dualvc,ibfvc,DBLP:conf/interspeech/YangDYZX22,fasts2svc, parrotron} with left-shifted convolution kernels that restrict its receptive field from accessing future frames. However, causal convolutions prevent models from fully utilizing within-chunk future context, causing performance degradation. To address this limitation, a solution termed dynamic chunk convolution is introduced in~\cite{dynamicconv} for streaming ASR. This technique involves training the model with chunked input and using non-causal convolution without access to any future context beyond the current chunk's right boundary, preventing mode mismatch between training and inference. 

During inference, future information is absent in the final convolutional receptive field of the current chunk, while available in the initial convolutional receptive field of the subsequent chunk. Neighboring convolutional input feature changes lead to an abrupt discontinuity in output features between these two chunks. Such feature discontinuities, while having minimal impact on ASR models due to their classification nature, can cause audible clicking sounds in the context of voice conversion. Therefore, directly employing dynamic chunk convolution for streaming voice conversion is not feasible.

To address this issue, we propose \textbf{dynamic masked convolution (DMC)} which is a novel dynamic masking strategy for convolutional inputs. The motivation is to enhance the robustness of the non-causal convolution to varying future information, allowing it to generate output features continuously without introducing clicking sounds between chunks. During the training process, within each convolution operation, the last $n$ frames which stand for future information in the kernel's receptive field are masked to zero, where:

\begin{equation}
n=rand(0,kernel/2).
\end{equation}

In a typical convolutional computation, the convolution kernel automatically moves over the input sequence to compute the complete output sequence, and we cannot apply different masks to the input for a single convolutional operation.
To overcome this limitation, as depicted in Fig.~\ref{fig:conv}, a functionally equivalent 2D convolution is employed to replicate the original 1D convolution process. This involves expanding the input sequence by adding an extra axis. The masking procedure can then be applied along this additional axis. By adopting this dynamic masked convolution technique, the streaming model can effectively take advantage of future information within chunks without clicking sounds between successive chunks.

\subsection{Quiet Attention}
In the conventional self-attention mechanism~\cite{attention}, the attention score matrix $W^{T\times T}$ is calculated using the softmax function:
\begin{equation}
\hat{w_{ti}} = Softmax(w_{ti}) = \frac{exp(w_{ti})}{\sum_{n=1}^{T}exp(w_{tn})},
\end{equation}
wherein the weights at time step $t$, denoted as $\hat{w_{ti}} \in W$, are normalized to sum up to one. 

Although the pre-trained ASR can remove most of the noise, there is still some remaining.
Therefore, our objective is to enhance the model's robustness against noise interference. 
In instances of noisy unvoiced frames at time $t$, it is crucial that attention calculations do not contribute any information. Even if all $w_{ti}$ tend towards negative infinity, the output probability $\hat{w_{ti}}$ is still computed to be $\frac{1}{T}$. To address this issue, we introduce the concept of ``quiet attention", as proposed in the reference\textsuperscript{\ref {quiet-attention}}. The quiet attention mechanism can be defined as follows:

\begin{equation}
Softmax_1(w_{ti}) = \frac{exp(w_{ti})}{1 + \sum_{n=1}^{T}exp(w_{tn})},
\end{equation}
introducing an escape mechanism in the negative orthant. Incorporating quiet attention allows us to ignore any information coming from unvoiced frames, thereby eliminating residual noise and preventing clutter.

\subsection{Data Augmentation}
Compared to the clean speech clips present in the training dataset, the recording environment for actual speech is notably more intricate, characterized by the presence of background noise and reverberation. Moreover, the speaking style found within the training data mainly involves scripted reading, while real-life conversations occur at a significantly swifter pace than the training data, consequently giving rise to challenges in maintaining model intelligibility.

To address this issue, we adopt data augmentations. Initially, we employ noise augmentation utilizing the MUSAN noise dataset~\cite{musan} through direct addition. Subsequently, we introduce random reverberation and tempo augmentations. All of these augmentations are done using the open-source tool WavAugment\footnote{https://github.com/facebookresearch/WavAugment/}.

\section{Experiments}
\begin{table*}[!htbp]
\caption{Comparison of the proposed DualVC 2 with DualVC, IBF-VC, and ablation models regarding speaker similarity and speaker naturalness MOS with confidence intervals of 95\% under 2 voice conversion scenarios. NMOS denotes naturalness MOS, and SMOS denotes speaker similarity MOS. A higher value means better performance.}
\vspace{-8pt}
 \label{tab:mos}
\setlength{\tabcolsep}{3mm}
 \centering
 \resizebox{\linewidth}{!}{
 \renewcommand{\arraystretch}{1.2}
\begin{tabular}{l|ccc|ccc|ccc}
\hline
           & \multicolumn{3}{c|}{Clean}                           & \multicolumn{3}{c|}{Noisy}                           & \multicolumn{3}{c}{Overall}                          \\ \hline
           & NMOS $\uparrow$ & SMOS $\uparrow$ & CER(\%) $\downarrow$ & NMOS $\uparrow$ & SMOS $\uparrow$ & CER(\%) $\downarrow$ & NMOS $\uparrow$ & SMOS $\uparrow$ & CER(\%) $\downarrow$ \\ \hline
DualVC (non-streaming)             & 3.88$\pm$0.02   & 3.80$\pm$0.03   & 8.9                  & 3.72$\pm$0.05   & 3.74$\pm$0.02   & 10.1                 & 3.80$\pm$0.02   & 3.77$\pm$0.04   & 9.5                  \\
DualVC (streaming)                 & 3.73$\pm$0.04   & 3.75$\pm$0.05  & 10.4                 & 3.64$\pm$0.03   & 3.68$\pm$0.04   & 12.5                 & 3.69$\pm$0.03   & 3.72$\pm$0.02   & 11.5                  \\
IBF-VC                             & 3.66$\pm$0.04   & 3.76$\pm$0.02   & 13.3                 & 3.58$\pm$0.02   & 3.72$\pm$0.02   & 15.6                 & 3.62$\pm$0.02   & 3.74$\pm$0.03   & 14.5                 \\ \hline
DualVC 2 (non-streaming)           & 4.07$\pm$0.03   & 3.90$\pm$0.04   & 8.3                  & 3.89$\pm$0.03   & 3.80$\pm$0.03   & 9.7                  & 3.98$\pm$0.04   & 3.85$\pm$0.03   & 9.0                  \\
DualVC 2 (streaming)               & 3.89$\pm$0.05   & 3.83$\pm$0.04   & 9.9                  & 3.78$\pm$0.03   & 3.75$\pm$0.02   & 10.8                 & 3.84$\pm$0.03   & 3.79$\pm$0.04   & 10.4                  \\
\hspace{1em}-DMC                   & 3.32$\pm$0.02   & 3.67$\pm$0.05   & 15.6                 & 3.24$\pm$0.05   & 3.60$\pm$0.02   & 18.3                 & 3.28$\pm$0.04   & 3.64$\pm$0.02   & 17.0                 \\
\hspace{1em}-Quiet Attention       & 3.82$\pm$0.04   & 3.76$\pm$0.02   & 10.2                 & 3.63$\pm$0.04   & 3.73$\pm$0.05   & 15.3                 & 3.72$\pm$0.02   & 3.75$\pm$0.04   & 12.8                 \\
\hspace{1em}-Data Aug              & 3.78$\pm$0.02   & 3.74$\pm$0.04   & 12.0                 & 3.52$\pm$0.02   & 3.67$\pm$0.05   & 16.1                 & 3.65$\pm$0.03   & 3.71$\pm$0.02   & 14.1                 \\ \hline
\end{tabular}
}
\vspace{-12pt}
\end{table*}
In the experiments, all testing VC models are trained on an open-source Mandarin corpus AISHELL-3~\cite{aishell3}. This dataset encompasses 88,035 utterances spoken by 218 speakers. Among these, a male and a female speaker are reserved as target speakers for evaluation. 100 clean and 100 noisy clips are used as source recordings. The selected recordings are then converted to the two target speakers using the proposed model and all comparison models to further perform evaluations. 

All the speech utterances are resampled to 16 kHz for VC training. Mel-spectrogram and BNF are computed at a frame length of 50 ms and a frame shift of 12.5 ms. The ASR encoder for BNF extraction is a 5-layer Conformer that comes from Fast-U2++ ~\cite{fastu2++} which is implemented by WeNet toolkit~\cite{wenet}, and trained on a Mandarin ASR corpus WenetSpeech~\cite{wenetspeech}. 
Bottleneck features are extracted from the last layer of the ASR encoder.
The speaker embedding is extracted using WeSpeaker toolkit~\cite{wespeaker}. To reconstruct waveform from the converted Mel-spectrograms, we use HiFi-GAN ~\cite{hifigan} with iSTFT upsampling layers~\cite{istftnet} to perform high-fidelity while fast waveform generation. It generates 24 kHz waveforms from 16 kHz spectrograms for better sound quality.

DualVC 2 consists of 6 Conformer blocks with 256 feature dimensions and 4 self-attention heads, the encoder and decoder each having 3 layers. The future prediction step is 6 for the HPC module. To evaluate the performance of the proposed model, \textbf{DualVC}~\cite{dualvc} in our previous work along with reimplemented \textbf{IBF-VC} ~\cite{ibfvc} are selected as baseline systems.

\vspace{-4pt}
\subsection{Subjective Evaluation}
\vspace{-4pt}
We conduct Mean Opinion Score (MOS) tests to evaluate the naturalness and speaker similarity of comparison models. The naturalness metric mainly considers intelligibility, prosody, and sound quality. A higher naturalness MOS score indicates the converted speech sounds more human-like. The similarity test uses the target speaker's real recording as the reference to evaluate the speaker timbre similarity between real and converted recordings. In both MOS tests, there are 30 listeners participated. 
\vspace{-4pt}

\subsubsection{Speech Naturalness}
\vspace{-4pt}
The naturalness MOS results presented in Table~\ref{tab:mos} indicate that our proposed DualVC 2 can achieve the best performance in speech naturalness. Notably, when fed with clean input, the streaming version of DualVC 2 surpasses IBF-VC and even outperforms non-streaming DualVC in terms of MOS scores. Compared to other baseline systems, DualVC 2 has the least performance degradation with noise input, proving its superior robustness. With the dynamic masked convolution effectively capturing within-chunk future information, streaming DualVC 2 can achieve performance close to its non-streaming mode.
\vspace{-4pt}
\subsubsection{Speaker Similarity}
\vspace{-4pt}
The results of similarity MOS tests among comparison models are also shown in Table 1. Unlike the naturalness metrics, the speaker similarity scores across different models are closer, reflecting the excellent decoupling ability of the recognition-synthesis framework. In terms of relative scores, the overall trend is close to the naturalness MOS, with DualVC 2 getting the highest scores while being robust under noisy inputs.
Considering both speaker similarity and naturalness performance, DualVC 2 exhibits a remarkable superiority for streaming voice conversion.

\subsubsection{Ablation Study}
To investigate the importance of our proposed methods in DualVC 2, three ablation systems were obtained by dropping dynamic masked convolution (-DMC), quiet attention (-Quiet Attention), and data augmentation (-Data Aug). 
As shown in Table 1, the removal of these methods brings obvious performance declines with respect to both speech naturalness and speaker similarity. 
Notably, the elimination of the dynamic masked convolution brings the most performance decline, the obvious clicking sound makes the converted result significantly less natural. This observation demonstrates that the utilization of future context within chunks is a significant boost to model performance. The removal of quiet attention and data augmentation has little effect on clean data, but the naturalness decreases significantly on noisy data, reflecting the enhancement of model noise robustness by these two methods.

\begin{figure}
\centering
\subfloat[DualVC 2]{%
  \includegraphics[width=2.6cm]{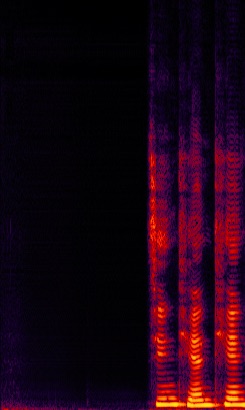}%
  \label{noisy}%
}\quad
\subfloat[w/o Quiet Attention]{%
  \includegraphics[width=2.6cm]{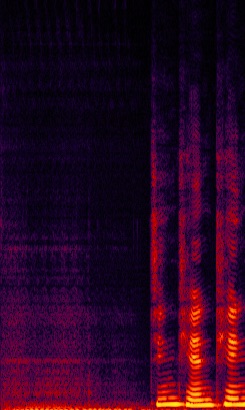}%
}\quad
\subfloat[w/o DMC]{%
  \includegraphics[width=2.6cm]{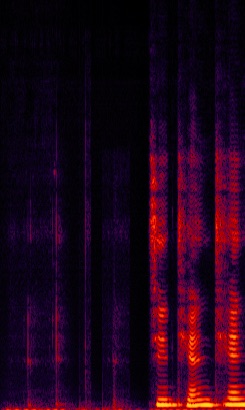}%
}
\caption{Visualizations of generated mel-spectrograms by DualVC 2 and ablation systems.}
\vspace{-12pt}
\label{fig:spec}
\end{figure}

We also examined spectrograms generated by the proposed DualVC 2, without quiet attention and without DMC. 
It can be seen in Fig~\ref{fig:spec} that noise is preserved in unvoiced frames by removing quiet attention, while removing the DMC introduces vertical lines in the spectrogram, resulting in audible clicking sounds.

\subsection{Objective Evaluation}

\subsubsection{Intelligibility Evaluation}
We employ the same pre-trained ASR to extract BNFs and transcribe the source and VC-generated speech clips. In order to ensure the accuracy of our results, we conduct testing on a larger dataset comprising 500 samples. The Character Error Rate (CER) is also detailed in Table~\ref{tab:mos}. For the source speech, we observed a CER of 6.6\% for clean clips and 8.6\% for noisy ones.
Streaming DualVC 2 induces a small CER increase compared to its non-streaming version, while both outperform baseline systems, demonstrating the ability to achieve good intelligibility in both clean and noisy scenarios.

\subsubsection{Computational Efficiency Evaluation}
In this study, we assess computational efficiency using three key metrics: real-time factor (RTF), latency, and parameters, as summarized in Table ~\ref{tab:performance}. RTF is a widely used measure for evaluating model inference speed, representing the ratio between model inference time and input feature duration. To satisfy real-time requirements, the RTF should be less than 1, and our complete pipeline achieved an impressive RTF of 0.165 when running on a single Intel i5-10210U core. Latency, on the other hand, is defined as the time interval from user input to model output, encompassing three components: model inference, input waiting, and network latency. Excluding network latency, system latency can be expressed as:
\begin{equation}
Latency = chunksize \times (1 + RTF).
\end{equation}
With a chunk size of 160 ms and a model inference latency of 26.4 ms, the total pipeline latency was calculated to be 186.4 ms. The parameter of all three models is 29 M. Compared to DualVC's RTF of 0.58 and latency with 252.8 ms at 41M parameters, DualVC 2 has a huge improvement in computational performance.

\begin{table}[]
\centering
 \caption{Real-time \& Computation Metrics of DualVC 2 tested on single core i5 10210U. Latency is obtained by multiplying the RTF by the chunk size.}
 \vspace{-8pt}
\setlength{\tabcolsep}{0.6mm}
 \label{tab:performance}
\setlength{\tabcolsep}{3mm}{
\resizebox{0.9\linewidth}{!}{
\begin{tabular}{l|ccc}
\hline
        & RTF           & Latency (ms)      & Params (M)                \\ \hline
ASR encoder     & 0.038         & 6.08      & 20.3                 \\
DualVC 2  & 0.083          & 13.28        & 7.5              \\
Vocoder & 0.044         & 7.04            & 1.2                 \\ \hline
All     & 0.165        & 26.40            & 29.0                   \\ \hline
\end{tabular}
}
}
\vspace{-12pt}
\end{table}

\section{Conclusions}

\label{sec:typestyle}
In this work, we upgrade our previous dual-mode voice conversion system DualVC to its new version DualVC 2. Built on the recognition-synthesis framework, DualVC 2 uses Conformer as the backbone for its excellent contextual information extraction and parallel computing capabilities. To better leverage future information within chunks, we propose dynamic masked convolution to make non-causal convolution applicable to streaming inference. We take advantage of quiet attention along with data augmentation to enhance the robustness of DualVC 2. Experiments show that DualVC 2 outperforms the baseline systems with an RTF of only 0.165 and a pipeline latency of 186.4 ms.

\vfill
\pagebreak


\bibliographystyle{IEEEbib}
\bibliography{strings,refs}

\begin{thebibliography}{10}

\bibitem{vcoverview}
Berrak Sisman, Junichi Yamagishi, Simon King, and Haizhou Li,
\newblock ``An overview of voice conversion and its challenges: From
  statistical modeling to deep learning,''
\newblock {\em {IEEE} {ACM} Trans. Audio Speech Lang. Process.}, vol. 29, pp.
  132--157, 2021.

\bibitem{bgmvc}
Jixun Yao, Yi~Lei, Qing Wang, Pengcheng Guo, Ziqian Ning, Lei Xie, Hai Li,
  Junhui Liu, and Danming Xie,
\newblock ``Preserving background sound in noise-robust voice conversion via
  multi-task learning,''
\newblock in {\em Proc. ICASSP}. 2023, pp. 1--5, {IEEE}.

\bibitem{expressive-vc}
Ziqian Ning, Qicong Xie, Pengcheng Zhu, Zhichao Wang, Liumeng Xue, Jixun Yao,
  Lei Xie, and Mengxiao Bi,
\newblock ``Expressive-vc: Highly expressive voice conversion with attention
  fusion of bottleneck and perturbation features,''
\newblock in {\em Proc. ICASSP}. 2023, pp. 1--5, {IEEE}.

\bibitem{voiceprivacy2022}
Jixun Yao, Qing Wang, Li~Zhang, Pengcheng Guo, Yuhao Liang, and Lei Xie,
\newblock ``{NWPU-ASLP} system for the voiceprivacy 2022 challenge,''
\newblock {\em CoRR}, vol. abs/2209.11969, 2022.

\bibitem{ELECTROLARYNGEAL}
Kazuhiro Kobayashi, Tomoki Hayashi, and Tomoki Toda,
\newblock ``Low-latency electrolaryngeal speech enhancement based on
  fastspeech2-based voice conversion and self-supervised speech
  representation,''
\newblock in {\em Proc. ICASSP}, 2023, pp. 1--5.

\bibitem{acevc}
Shehzeen Hussain, Paarth Neekhara, Jocelyn Huang, Jason Li, and Boris Ginsburg,
\newblock ``{ACE-VC:} adaptive and controllable voice conversion using
  explicitly disentangled self-supervised speech representations,''
\newblock {\em CoRR}, vol. abs/2302.08137, 2023.

\bibitem{vqmivc}
Disong Wang, Liqun Deng, Yu~Ting Yeung, Xiao Chen, Xunying Liu, and Helen Meng,
\newblock ``{VQMIVC:} vector quantization and mutual information-based
  unsupervised speech representation disentanglement for one-shot voice
  conversion,''
\newblock in {\em Proc. INTERSPEECH}. 2021, pp. 1344--1348, {ISCA}.

\bibitem{stylevc}
Zhichao Wang, Xinyong Zhou, Fengyu Yang, Tao Li, Hongqiang Du, Lei Xie, Wendong
  Gan, Haitao Chen, and Hai Li,
\newblock ``Enriching source style transfer in recognition-synthesis based
  non-parallel voice conversion,''
\newblock in {\em Proc. INTERSPEECH}. 2021, pp. 831--835, {ISCA}.

\bibitem{dualvc}
Ziqian Ning, Yuepeng Jiang, Pengcheng Zhu, Jixun Yao, Shuai Wang, Lei Xie, and
  Mengxiao Bi,
\newblock ``Dualvc: Dual-mode voice conversion using intra-model knowledge
  distillation and hybrid predictive coding,''
\newblock in {\em Proc. INTERSPEECH}. 2023, pp. 2063--2067, {ISCA}.

\bibitem{ibfvc}
Yuanzhe Chen, Ming Tu, Tang Li, Xin Li, Qiuqiang Kong, Jiaxin Li, Zhichao Wang,
  Qiao Tian, Yuping Wang, and Yuxuan Wang,
\newblock ``Streaming voice conversion via intermediate bottleneck features and
  non-streaming teacher guidance,''
\newblock {\em CoRR}, vol. abs/2210.15158, 2022.

\bibitem{DBLP:conf/interspeech/YangDYZX22}
Haoquan Yang, Liqun Deng, Yu~Ting Yeung, Nianzu Zheng, and Yong Xu,
\newblock ``Streamable speech representation disentanglement and multi-level
  prosody modeling for live one-shot voice conversion,''
\newblock in {\em Proc. INTERSPEECH}. 2022, {ISCA}.

\bibitem{fasts2svc}
Hirokazu Kameoka, Kou Tanaka, and Takuhiro Kaneko,
\newblock ``Fasts2s-vc: Streaming non-autoregressive sequence-to-sequence voice
  conversion,''
\newblock {\em CoRR}, vol. abs/2104.06900, 2021.

\bibitem{parrotron}
Oleg Rybakov, Fadi Biadsy, Xia Zhang, Liyang Jiang, Phoenix Meadowlark, and
  Shivani Agrawal,
\newblock ``Streaming parrotron for on-device speech-to-speech conversion,''
\newblock {\em CoRR}, vol. abs/2210.13761, 2022.

\bibitem{cpc}
A{\"{a}}ron van~den Oord, Yazhe Li, and Oriol Vinyals,
\newblock ``Representation learning with contrastive predictive coding,''
\newblock {\em CoRR}, vol. abs/1807.03748, 2018.

\bibitem{apc}
Yu{-}An Chung, Wei{-}Ning Hsu, Hao Tang, and James~R. Glass,
\newblock ``An unsupervised autoregressive model for speech representation
  learning,''
\newblock in {\em Proc. INTERSPEECH}. 2019, pp. 146--150, {ISCA}.

\bibitem{recognition-synthesis}
Jing{-}Xuan Zhang, Zhen{-}Hua Ling, and Li{-}Rong Dai,
\newblock ``Non-parallel sequence-to-sequence voice conversion with
  disentangled linguistic and speaker representations,''
\newblock {\em {IEEE} {ACM} Trans. Audio Speech Lang. Process.}, vol. 28, pp.
  540--552, 2020.

\bibitem{Conformer}
Anmol Gulati, James Qin, Chung{-}Cheng Chiu, Niki Parmar, Yu~Zhang, Jiahui Yu,
  Wei Han, Shibo Wang, Zhengdong Zhang, Yonghui Wu, and Ruoming Pang,
\newblock ``Conformer: Convolution-augmented transformer for speech
  recognition,''
\newblock in {\em Proc. INTERSPEECH}. 2020, pp. 5036--5040, {ISCA}.

\bibitem{wenet}
Zhuoyuan Yao, Di~Wu, Xiong Wang, Binbin Zhang, Fan Yu, Chao Yang, Zhendong
  Peng, Xiaoyu Chen, Lei Xie, and Xin Lei,
\newblock ``Wenet: Production oriented streaming and non-streaming end-to-end
  speech recognition toolkit,''
\newblock in {\em Proc. INTERSPEECH}. 2021, pp. 4054--4058, {ISCA}.

\bibitem{dynamicconv}
Xilai Li, Goeric Huybrechts, Srikanth Ronanki, Jeff Farris, and Sravan
  Bodapati,
\newblock ``Dynamic chunk convolution for unified streaming and non-streaming
  conformer asr,''
\newblock in {\em Proc. ICASSP}. IEEE, 2023, pp. 1--5.

\bibitem{attention}
Ashish Vaswani, Noam Shazeer, Niki Parmar, Jakob Uszkoreit, Llion Jones,
  Aidan~N. Gomez, Lukasz Kaiser, and Illia Polosukhin,
\newblock ``Attention is all you need,''
\newblock in {\em Proc. NeurIPS}, 2017, pp. 5998--6008.

\bibitem{musan}
David Snyder, Guoguo Chen, and Daniel Povey,
\newblock ``{MUSAN:} {A} music, speech, and noise corpus,''
\newblock {\em CoRR}, vol. abs/1510.08484, 2015.

\bibitem{aishell3}
Yao Shi, Hui Bu, Xin Xu, Shaoji Zhang, and Ming Li,
\newblock ``{AISHELL-3:} {A} multi-speaker mandarin {TTS} corpus,''
\newblock in {\em Proc. INTERSPEECH}. 2021, pp. 2756--2760, {ISCA}.

\bibitem{fastu2++}
Chengdong Liang, Xiao{-}Lei Zhang, Binbin Zhang, Di~Wu, Shengqiang Li, Xingchen
  Song, Zhendong Peng, and Fuping Pan,
\newblock ``Fast-u2++: Fast and accurate end-to-end speech recognition in joint
  ctc/attention frames,''
\newblock {\em CoRR}, vol. abs/2211.00941, 2022.

\bibitem{wenetspeech}
Binbin Zhang, Hang Lv, Pengcheng Guo, Qijie Shao, Chao Yang, Lei Xie, Xin Xu,
  Hui Bu, Xiaoyu Chen, Chenchen Zeng, Di~Wu, and Zhendong Peng,
\newblock ``{WENETSPEECH:} {A} 10000+ hours multi-domain mandarin corpus for
  speech recognition,''
\newblock in {\em Proc. ICASSP}. 2022, pp. 6182--6186, {IEEE}.

\bibitem{wespeaker}
Hongji Wang, Chengdong Liang, Shuai Wang, Zhengyang Chen, Binbin Zhang,
  Xu~Xiang, Yanlei Deng, and Yanmin Qian,
\newblock ``Wespeaker: {A} research and production oriented speaker embedding
  learning toolkit,''
\newblock {\em CoRR}, vol. abs/2210.17016, 2022.

\bibitem{hifigan}
Jiaqi Su, Zeyu Jin, and Adam Finkelstein,
\newblock ``Hifi-gan: High-fidelity denoising and dereverberation based on
  speech deep features in adversarial networks,''
\newblock in {\em Proc. INTERSPEECH}, Helen Meng, Bo~Xu, and Thomas~Fang Zheng,
  Eds. 2020, pp. 4506--4510, {ISCA}.

\bibitem{istftnet}
Takuhiro Kaneko, Kou Tanaka, Hirokazu Kameoka, and Shogo Seki,
\newblock ``{ISTFTNET:} fast and lightweight mel-spectrogram vocoder
  incorporating inverse short-time fourier transform,''
\newblock in {\em Proc. ICASSP}. 2022, pp. 6207--6211, {IEEE}.

\end{thebibliography}

\end{document}